# On the Non-observability of Recent Biogenesis

CHARLES H. LINEWEAVER[1,2] and TAMARA M. DAVIS[1]
[1] School of Physics, University of New South Wales, Sydney, Australia
[2] Australian Centre for Astrobiology, Macquarie University, Sydney, Australia
charley@bat.phys.unsw.edu.au

In our article "Does the rapid appearance of life on Earth suggest that life is common in the Universe?" (Lineweaver & Davis 2003, hereafter LD) we showed how the probability of biogenesis on Earth-like planets in the Universe could be inferred from the rapidity of biogenesis on the only example we know of: Earth. We argued that although nothing could be inferred from the fact that biogenesis had happened on our planet, interesting constraints could be inferred from the rapidity of that biogenesis. We found that the probability of biogenesis on Earth-like planets in the Universe was close to unity and that at the 95% confidence level, more than 13% of Earth-like planets would have had biogenesis (conditions apply).

The most critical condition upon which this result depended was that rapid biogenesis was not a requirement for our existence. "Any effect that makes rapid biogenesis a prerequisite for life would undermine our inferences for q." (LD Section 3.3). If it were the case that life, if it is to evolve at all, must do so early in the history of an Earth-like planet, then our argument would fall apart. Rapid biogenesis would then be a requirement for our existence. All life forms in the Universe (no matter how common or even infinitesiamally rare) would look back at their own history and find that life evolved rapidly on their planets. For example, suppose the detailed but unknown steps of molecular evolution that lead to the origin of life were not frustrated by, but required a thermal and molecular environment that could only be found at the end of the late heavy bombardment 4.0 to 3.8 billion years ago. In this case the inferences in LD would be invalid -- even if we have good evidence to suspect that the same thermal and molecular environments and late heavy bombardments can be found in the early history of terrestrial planets everywhere in the Universe.

In the accompanying article Flambaum argues that there are "a large number of crucial steps between the first live organisms and humans". These steps take a long time. Therefore for us to be here now, the initial step, biogenesis, had to be rapid (to allow enough time for the large number of crucial steps to be completed). In our paper we called this selection effect the "non-observability of recent biogenesis" and discussed it in Section 3.2. We agree with Flambaum that this selection effect means that biogenesis must be non-recent. However, in contrast to Flambaum, we argued that "non-recent" does not necessarily imply "rapid". The distinction is important to this discussion. "The selection effect for non-recent biogenesis is selecting for biogenesis to happen a few billion years before the present regardless of whether it happened rapidly. It is not a selection effect for rapid biogenesis since the longer it took us to evolve to a point when we could measure the age of the Earth, the older the Earth became. Similarly, if biogenesis took 1 Gyr longer than it actually did, we would currently find the age of the Earth to be 5.566 Gyr (= 4.566 + 1) old…" (LD Section 3.2).

Apparently Flambaum was not convinced by our argument. Nor were we entirely convinced. That is why we quantified the effect in Figures 4 and 5 (LD) and derived the probability of biogenesis as a function of the rather controversial unknown : How much longer could biogenesis have taken without diminishing severely the chances of evolving into observerhood? In the language of Fig. 5 (LD), Flambaum is arguing that N =1, i.e., that biogenesis could not have taken longer than it did because there had to be sufficient time left after biogenesis (~ 3.5 Gyr in our case) to complete the large number of crucial steps between the first organisms and humans. As Fig. 5 shows, if N=1 little can be inferred from rapid terrestrial biogenesis. In LD we accounted for this self-selection effect by minimizing the time we assumed that biogenesis could have taken. We assumed N=2 for the "13%" result quoted in the abstract.

The Sun is 4.56 Gyr old. Its total main sequence lifetime will be ~ 10 Gyr. Thus, it has another ~ 5 Gyr to go. We seem to have evolved with plenty of time left to enjoy our observerhood. The increasing luminosity of the Sun with age may shorten the time available, but according to our best estimates, we still have "at least another 1 Gyr and possibly much longer"(Caldiera & Kasting 1992). Let us call this available time still left to us $\Delta t_{left}$ and estimate it at $\Delta t_{left}$ ~ 2 +/- 1 Gyr. If the evolution of observers is so hard, why did we wake up to find that we still have ~ 2 Gyr left?

In an articulate paper, Hanson (1998) addressed this specific issue: "Must Early Life Be Easy?" He came to the following conclusions. In a sample that has successfully passed through a number of steps of varying difficulty ,
- the time intervals taken by the easiest steps reflect the probability of those steps while the time intervals taken by the most difficult steps do not.
- the time left over after completion of all the steps is approximately the same as the time taken by the hardest step.

Let $\Delta t_{biogenesis}$ be how long biogenesis took on Earth. According to Hanson's analysis, if biogenesis had been a hard step we would expect $\Delta t_{left}$ ~ $\Delta t_{biogenesis}$. However, this does not seem to be the case. With $\Delta t_{left}$ ~ 2 +/- 1 Gyr (Caldiera & Kasting 1992) and $\Delta t_{biogenesis}$ = $0.1^{+0.5}_{-0.1}$ (LD Section 2) we have $\Delta t_{left}$ ~ 20 $\Delta t_{biogenesis}$ . This suggests that biogenesis was not one of the harder steps but was one of the easier ones -- that biogenesis did not have to be as rapid as it was, and therefore that its rapidity is a measure of its probability. However, within the error bars this factor of 20 could be as small as 2 or as large as a few million. More realistic modeling of the future of the biosphere may reduce the error bars on $\Delta t_{left}$ but the largest hope for progress will be in discoveries of earlier dates for the first life on Earth (or on Mars). Although very approximate, we interpret these numbers and this new $\Delta t_{left}$ constraint to provide marginal new evidence in favor of the idea that rapid biogenesis was not a requirement, thus providing support for the main conclusions of LD.

The use of the observational constraints on $\Delta t_{left}$ are relevant to Flambaum's argument. If the number of crucial steps n is much greater than one, as Flambaum assumes, and the amount of time taken to finish these steps is t, then based on Hanson's analysis, $\Delta t_{left}$ should satisfy $\Delta t_{left}$ ~ t/n, in contradiction to the observation that $\Delta t_{left}$ ~ t/2.

We do share Flambaum's motivating assumption that the evolution of humans (or any particular species) is unlikely or even a set of measure zero (see Simpson 1964). The evolutionary history of life on Earth strongly suggests that once extinct, species do not come back. Whether the more generic evolution of life forms worthy of being called "observers" is also a set of measure zero, is an important unresolved problem. If individual species and the other products of evolution are sets of measure zero, then ideas about the probability of "crucial steps" are inappropriate -- if you are not going anywhere, then no steps are crucial. Our inferences in LD are at least partially immune to this inappropriateness in the sense that the transition from the abiotic to the biotic may be a more deterministic process than the subsequent quirky products of biological evolution.